**CVD nanodiamonds with non-blinking, near transform-limited linewidths emitters**

Ke Li[1], Yu Zhou[1], A. Rasmita[1], I. Aharonovich[2, *], W. B. Gao[1, *]

1. Division of Physics and Applied Physics, School of Physical and Mathematical Sciences, Nanyang Technological University, Singapore 637371, Singapore
2. School of Mathematical and Physical Sciences, University of Technology Sydney, Ultimo, NSW, 2007, Australia
Email: Igor.Aharonovich@uts.edu.au, wbgao@ntu.edu.sg

**Abstract**
Near transform-limited single photon sources are required for perfect photon indistinguishability in quantum networks. Having such sources in nanodiamonds is particularly important since it can enable engineering hybrid quantum photonic systems. In this letter, we report the generation of optically stable, nearly transform-limited single silicon vacancy emitters in nanodiamonds. Lines as narrow as 325 MHz are reported, which is comparable to the lifetime limited linewidth. Moreover, the emitters exhibit minimal spectrum diffusion and absolute photostability, even if pumped well above saturation. Our results suggest that nanodiamonds can host color centers with supreme properties suitable for hybrid photonic devices and quantum information.

## I. Introduction

Fluorescent nanodiamonds that host bright and narrowband optical defects are attractive for numerous applications spanning quantum photonics[1-3], sensing[4-7] and bio-imaging[8-10]. One of these defects is the negatively charged silicon vacancy, (SiV$^-$), that has been subject to an intense research efforts due to its promising optical properties[11-19]. The defect consists of an interstitial silicon atom splitting two vacancies with a strong zero phonon line (ZPL) at 738 nm [20]. The structure of the defect is shown schematically in Fig. 1(a). It has a high Debye Waller factor, that results in ~ 70% of the emitted photons being into the ZPL, and therefore is suitable for a variety of photonic applications. Due to its inverse symmetry, the defect is not susceptible for electric fields fluctuations, and therefore exhibit ultimate spectral characteristics[15].

For many applications, including quantum bio-imaging, sensing and hybrid quantum photonics, the use of nanodiamonds is preferred[21, 22]. This is because nanodiamonds can be easily manipulated to a position of choice, for instance onto a photonic resonator, to achieve Purcell enhancement[23]. So far, only the nitrogen vacancy (NV) centers and the chromium centers in nanodiamonds have been studied using fluorescence resonant excitation (PLE)[24-26] but the emitters' linewidths were always broadened by strain or significant spectral diffusion that resulted in majority of lines being broader than 1 GHz.

In this work, we report on nearly lifetime limited characteristics of the SiV defects in nanodiamonds grown using chemical vapor deposition (CVD) method. The studied SiV emiters exhibit linewidths as narrow as 325 MHz and absolute photostability even when excited above saturation. The grown nanodiamonds with such ideal properties will be excellent sources for variety of applications spanning quantum photonics, communications and quantum sensing.

## II. Experiment

The SiV⁻ defects were incorporated into the diamond nanocrystals during microwave plasma CVD growth from 4-6 nm detonation nanodiamond seeds. The growth conditions were hydrogen/methane ratio of 100:1 at 60 Torr, with a microwave power of 900 W. Under these conditions, well faceted nanodiamonds with sizes between 100 – 300 nm can be grown, as shown in Fig. 1(b).

The experimental setup used to characterize the nanodiamonds is shown in Fig. 1(c). The sample was mounted on a stage which is fixed on top of a stacked XYZ steppers (Attocube) with nanometer precision. Both the sample stage and the stacked XYZ steppers are placed in a cryostat, which is used to stabilize the sample temperature at 4.2 K. A home-built confocal microscope is used to focus the excitation laser beam onto the diamond surface through a microscope objective (numerical aperture (NA) 0.80, X100) which is also used to collect the fluorescence. A pair of polarizer is used for cross-polarization rejection of exciting laser light. A multimode fiber (50 μm core diameter, NA 0.22) is used to collect the signal. The collected signal is guided into a spectrometer with its 1800 grating/mm or towards two avalanche photo diodes for antibunching measurments.

To measure ensemble photoluminescence (Fig. 2a), 690 nm laser is used for excitation and a 700 nm long pass filter was placed on the collection arm to block residual 690 nm excitation light. For photoluminescence excitation (PLE) experiment, we use Titanium:Sapphire (Ti:SAP) laser with less than 5 MHz linewidth and a wide tunable wavelength range between 700 - 800 nm. The PLE scan experiment is performed through scanning a Piezo Etalon within a mode hop free range of 50 GHz and detect the off-resonant phonon sideband after a 750 nm long pass filter. A wavemeter was used to monitor the real-time wavelength of exciting laser, as the feedback for precisely lock laser wavelength at specific value. With this method, laser wavelength can be locked at the peak of ZPL line of the single emitter, for the $g^2(\tau)$ and PLE stability measurements. Finally, lifetime measurements were recorded using a pulsed Ti:SAP laser with pulse length of ~10 ps.

## III. Results and Discussions

The measured spectrum of the SiV⁻ ensemble at 4 K is shown in Fig. 2(a). The ensemble represents different emitters that present in the same nanocrystal. Since the confocal microscope has a finite resolution, we were not able to resolve individual emitters using the off-resonance excitation.

To probe individual emitters, we performed photoluminescence excitation (PLE). PLE measurements is an excellent enabler to probe individual emitters from an ensemble, since only the emitters resonating with an excitation source will be probed. Spectra of individual SiVs defects are shown in Fig. 2(b). The spectra were obtained by sweeping the Ti:SAP across a frequency range of the ensemble of SiVs as shown in Fig. 2(a). Each of the measured ZPLs has a linewidth less than 500 MHz while the minimum spectral separation between ZPLs of different SiV- centers is larger than 5 GHz, which means a single emitter can be spectrally isolated.

To prove that only single emitters are probed, we chose a particular emitter at 738.94984 nm (shown with red arrow in Fig. 2(a)) and recorded the second order autocorrelation measurement, $g^2(\tau)$. The laser wavelength was scanned across the ZPL of this emitter and the measured $g^2(\tau)$ curve is shown in Fig. 2(c). The observed dip at zero delay time, $g^2(0) = 0.014\pm0.006$, satisfies the criteria of single-photon emitter ($g^2(0) <0.5$). No background correction was performed and the data was fit to a three level model[13]. Low $g^{(2)}(0)$ values are highly important to generate clean single photon sources for quantum information[27]. The fluorescent lifetime of this defect is measured to be $T_1=1.126\pm0.005$ ns (141 MHz), as shown in Fig. 2(d).

Fig. 3(a) shows a high resolution PLE spectrum of the same emitter using a low excitation power of 267 nW (~ ¼ of saturation power, as will be described later). The full width at half maximum (FWHM) of the linewidth is only 325±30 MHz, fit using a Lorentzian function. The observed value $\Gamma_{ND}$= 325 ± 30 MHz is relatively close to the lifetime limited value of $1/2\pi T_1$ = 141MHz and is comparable with values reported for SiV emitters occurring in bulk diamond[14, 15, 19]. It is also interesting to note that unlike the SiV, the typical linewidth for the nitrogen vacancy centers in nanodiamonds is often significantly broader than the lifetime limited value ($\Gamma_{ND}$ ~ 1.2 GHz in nanodiamonds vs $\Gamma_L$ ~ 16 MHz lifetime limited)[24, 25], making the SiV in nanodiamonds an attractive single photon source.

To show the stability of the emitter, the same PLE scan was repeated for 18 times. Fig. 3(b) shows the ZPL line position at each scan. The standard frequency deviation from the ZPL resonance is only 24 MHz, which is smaller than one tenth of the measured FWHM spectrum linewidth of the SiV- center. Moreover, we did not observe line position drifting away from the averaged line position and the defect was stable for the duration of the measurements. We therefore conclude that spectral diffusion is negligible for these emitters, unlike the other color centers in nanodiamonds[25, 26]. The broadening of the PLE resonance is likely due to residual phonon broadening or dipole-dipole interaction with the nearby SiV emitters in the same nanodiamond. We further studied the emitters' stability under different excitation regimes – namely below saturation, at saturation and above saturation. The results are shown in Fig. 3(c). No blinking is observed, even for power exceeding 10 times the saturation power for the SiV. This emphasizes the remarkable photostability of the emitters, and the high quality of the host nanodiamond matrix. Combined with the nearly transform limited lifetime, and negligible spectral diffusion, the

SiV emitters in nanodiamonds can be ideal candidates for sensing, quantum information science and hybrid nanophotonics applications.

Figure 4 shows the counts and linewidth as a function of the excitation power. Resonant excitation is a very efficient method to drive the optical transition of individual emitters, hence power broadening of ZPL linewidth is not negligible when exciting power is close to saturation or beyond it. The experiment was performed through repeated PLE scan across a ZPL wavelength for different exciting power ranging from 40 nW to 44 µW while detecting the phonon sideband with the APD. From each scan data, the peak counts and FWHM linewidth data are obtained through Lorentzian fitting to the PLE scan data. The peak photon counting represents the value of the emitter when the laser is exactly on resonance. Fig. 4(a) shows the plot of excitation power vs emitters' intensity. The peak counts increases fast when increasing the resonant exciting power within a small range from 40 nW to 10 uW, and reaches a steady state when exciting power is increased and beyond this range. The data fits well using equation

$$I(P) = I_\infty \frac{P/P_s}{1+P/P_s} \qquad (1)$$

where $I(P)$ is the count rate, which depends on excitation power $P$. The two fitting parameters are saturation power $P_s$ and maximum count rate when excitation power approaches infinite. The fitting results a saturation power, $P_s$ = 987±53 nW for this single emitter, and a maximum count rate 35.3±0.5 Kcps.

The FWHM linewidth of the single emitter remains narrow for excitation powers below 1.2 µW, and increases rapidly at higher excitation powers. These results are important since higher photon flux can be generated using 1 mW excitation, while the emission linewidth will not be compromised. Fig. 4(b) shows the magnified region of the plot at lower excitation powers. FWHM data is fit well with a function of $\gamma_0 + \Gamma\sqrt{1+P/P_s}$, where $\gamma_0$ is the constant value 24MHz as we measured in spectrum diffusion, $P$ is the power used and $\Gamma$ is the linewidth. From the fitting, a minimum linewidth 312±6 MHz is obtained for the low excitation power limit.

## IV.     Summary

To summarize, we report on resonant excitation of single SiV defects in nanodiamonds. The emitters exhibit nearly transform limited linewidth of ~ 325 ±30 MHz at 4 K, only a factor of 2.3 broader that the lifetime limited values of ~ 141 MHz. Remarkably, minimal spectral diffusion and no blinking was observed, indicating the excellent properties of the color centers in the grown CVD nanodiamonds. While further studies to unveil the reason for the line broadening are important, our current results hold great promise to harness CVD grown nanodiamonds hosting SiVs for coupling to photonic cavities and be utilized for quantum information processing.

During the preparation of the manuscript, we became aware that a similar work has appeared on the preprint server (arxiv)[28]. Our results are very much comparable, both reporting nearly lifetime limited linewidth from nanodiamonds. The main difference stems from the fact that we have not observed blinking in the emitter's fluorescence while the work by Jantzen et al., reports severe blinking of the SiV. This may originate from the different methods by which the nanodiamonds were synthesized (HPHT[28] vs CVD [our work]).

**Acknowledgments**

We thank Kerem Bray and Russell Sandstrom for assistance with the CVD process. I. A. is the recipient of an Australian Research Council Discovery Early Career Research Award (Project Number DE130100592). Partial funding for this research was provided by the Air Force Office of Scientific Research, United States Air Force. W. B. Gao acknowledges the strong support from the Singapore National Research Foundation through a Singapore 2015 NRF fellowship grant (NRF-NRFF2015-03) and its Competitive Research Programme (CRP Award No. NRF-CRP14-2014-02), and a start-up grant (M4081441) from Nanyang Technological University.

**Figures with Caption**

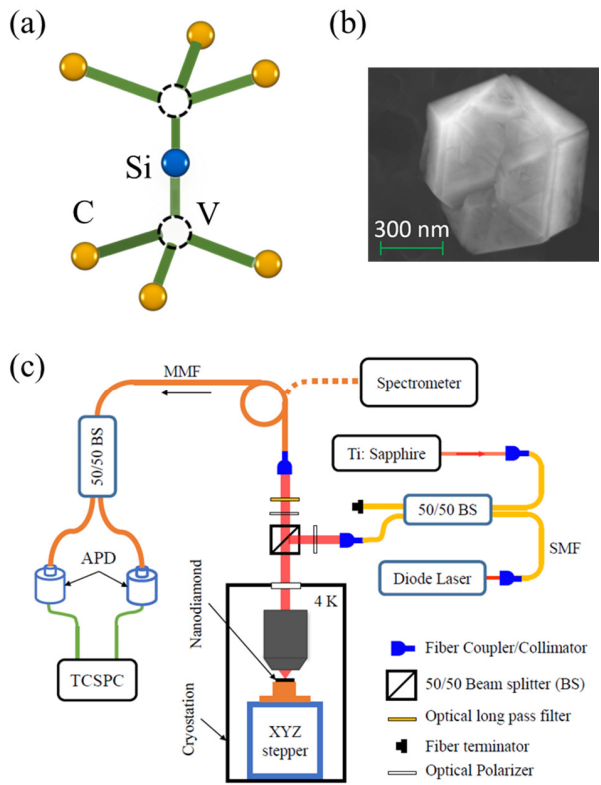

*Figure 1. (a) Schematic illustration of the silicon vacancy defect. An interstitial silicon atom splits two vacancies in a diamond lattice. (b) Scanning electron microscopy image of a typical nanodiamond. The clearly visible crystalline facets indicate on a high quality of the nanodiamond. (c) Experimental setup. 690 nm diode laser is used for PL measurement, Ti:Sapphire laser in cw mode is used for PLE excitation and g2 measurement. Pulsed Ti:Sapphire laser is used for lifetime measurement. For a detailed description, please refer to the main text.*

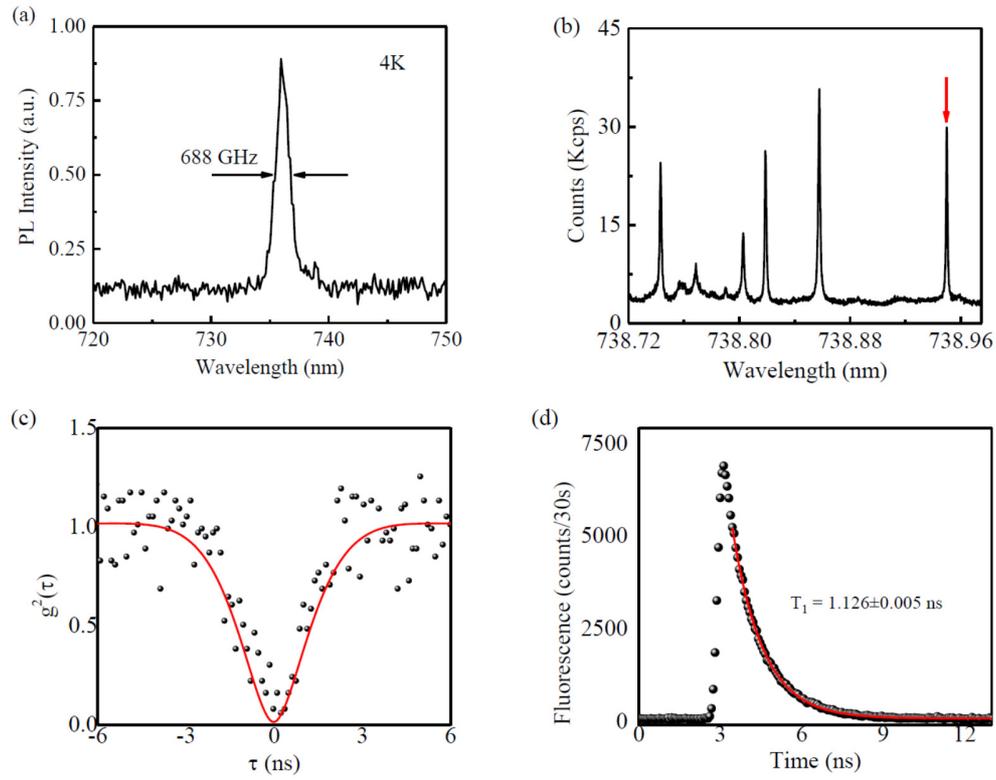

Figure 2: (a) off resonant photoluminescence spectrum from a single nanodiamond recorded at 4K using 690 nm excitation laser. (b) Resonance fluorescence spectrum of the sample showing several individual transitions of the SiV defects in a nanodiamond. Note that due to a close proximity of the SiVs in the nanodiamond, we are not able to absolutely specify which transitions are observed. (c) Second order correlation measurement, $g^{(2)}(\tau)$, demonstrating that a single SiV defect is addressed. The $g^{(2)}(\tau)$ curve is recorded from the 738.94983 nm line (red arrow in (b). Low excitation power (0.25 µW) is used to minimize the effect of power broadening. The phonon sideband is detected during $g^{(2)}(\tau)$ measurement. (d) Lifetime measurement of the same SiV center recorded using pulsed Ti:SAP.

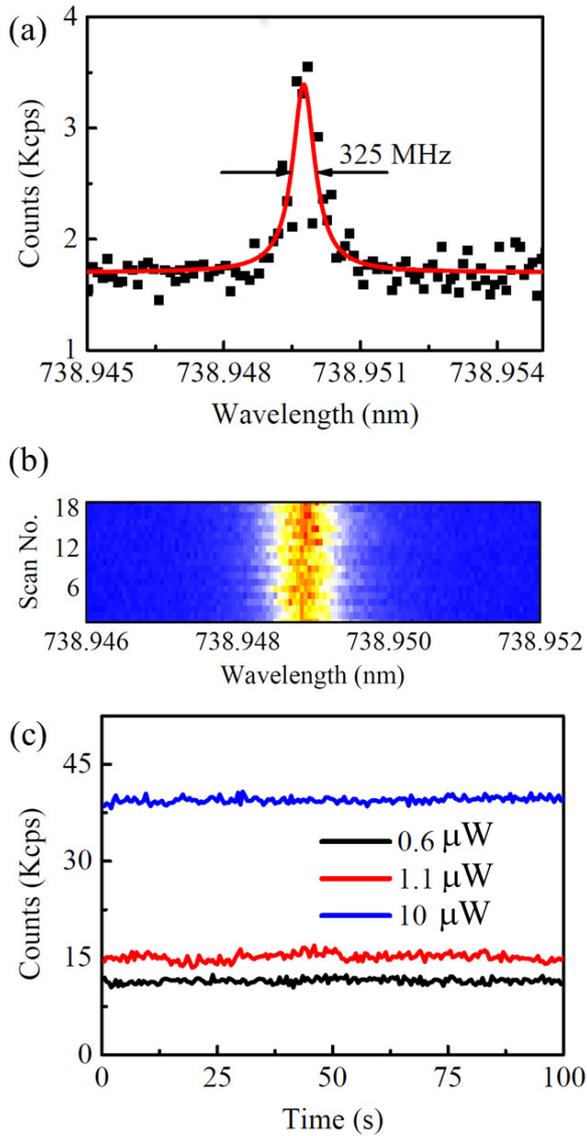

Figure 3. Photoluminescence Excitation (PLE) Spectroscopy study on a single emitter at 4K. The exciting laser wavelength is scanned through one of the resonant frequencies of one emitter while the PL from its phonon sideband (off-resonant) was detected by an APD. In order to minimize the power broadening, an exciting power well below saturation power is used for this experiment. (a) The 325 MHz FWHM linewidth is determined from Lorentz fitting to the PLE scan data curve. (b) 18 times PLE scan was done to examine the spectrum stability of the emitter. The ZPL line positions from 18 times PLE measurement distribute around a mean value 738.94983 nm with a standard deviation of 24 MHz. (c) The counts stability of the addressed SiV$^-$ centre is examined by locking the laser frequency to be on resonant with the defect, the monitored photon counts is always stable when exciting power was set at different value, 0.6 μW (<$P_s$), 1.1 μW (~$P_s$) and 10 μW (>> $P_s$)

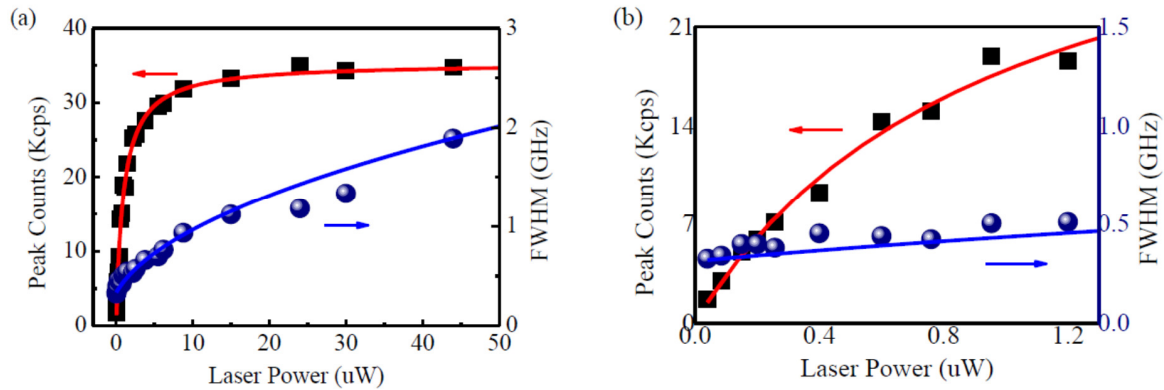

Figure 4. (a) Saturation curve of resonant photoluminescence of single SiV⁻ centre. Laser wavelength was scanned through ZPL of the single emitter while detecting PL of the phonon sideband with an APD. Lorentzian fit to each scan data is performed to extract the FHWH linewidth of ZPL and peak counts when laser is on resonant with the single emitter. The resonant photoluminescence from an individual SiV⁻ saturates fast when the resonant excitation power is increased from 40 nW to 10 μW. The saturation power and saturation counts obtained from fitting are 972±51 nW and 35.3±0.5 Kcps separately. ZPL linewidth (FWHM) of the single emitter approaches to a limit of 312± 6 MHz for zero exciting power, which is determined through fitting function $\gamma_0 + \Gamma\sqrt{1 + P/P_s}$, to FWHM data. (b) Zoomed-in area of lower excitation power regime where the FWHM is relatively constant.